\documentclass[%
 reprint,
 amsmath,amssymb,
 aps,
floatfix,
]{revtex4-1}
\usepackage{graphicx}
\usepackage{dcolumn}
\usepackage{bm}

\newcommand{\beginsupplement}{%
        \setcounter{table}{0}
        \renewcommand{\thetable}{S\arabic{table}}%
        \setcounter{figure}{0}
        \renewcommand{\thefigure}{S\arabic{figure}}%
     }

\begin{document}

\preprint{APS/123-QED}

\title{Spatial Coherence of a Strongly Interacting Bose Gas in the Trimerized Kagome Lattice}

\author{Thomas H. Barter$^1$}
\author{Tsz-Him Leung$^1$}
\author{Masayuki Okano$^1$}
\author{Maxwell Block$^1$}
\author{Norman Y. Yao$^{1,2}$}
\author{Dan M. Stamper-Kurn$^{1,2}$}

\affiliation{
    $^1$Department of Physics, University of California, Berkeley CA 94720 \\
    $^2$Materials Sciences Division, Lawrence Berkeley National Laboratory, Berkeley, CA 94720}

\date{\today}

\begin{abstract}

We produce a trimerized kagome lattice for ultracold atoms using an optical superlattice formed by overlaying  triangular lattices generated with two colors of light at a 2:1 wavelength ratio.  Adjusting the depth of each lattice tunes the strong intra-trimer ($J$) and weak inter-trimer ($J'$) tunneling energies, and also the on-site interaction energy $U$. Two different trimerization patterns are distinguished using matter-wave diffraction.  We characterize the coherence of a strongly interacting Bose gas in this lattice, observing persistent nearest-neighbor spatial coherence in the large $U/J'$ limit, and that such coherence displays asymmetry between the strongly and the weakly coupled bonds.
\end{abstract}

\maketitle

Ultracold atoms in optical lattices provide highly tunable implementations of condensed matter models. Various lattice geometries have been realized, including plaquette lattices, in which identical few-site plaquettes, with strong intra-plaquette coupling, are arrayed in a lattice and weakly coupled to one another.  Fine quantum control of the few-body states and dynamics within plaquettes have enabled experimental demonstrations of superexchange, resonant valence states, quantum magnetism, and anyonic statistics \cite{trot08superexchange, nasc12rvb, grei13, dai17toric}.

\begin{figure*}[t]
\includegraphics{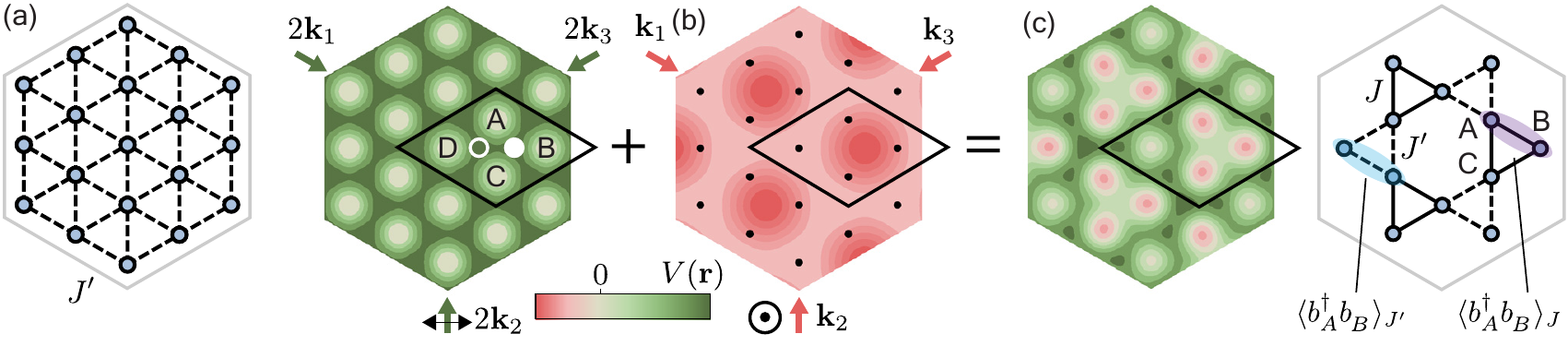}
 \caption{Construction of the optical trimerized kagome lattice.  (a) A SW triangular lattice of spacing $a_\triangle =355$ nm is formed by in-plane polarized 532-nm-wavelength light. Sites of a 2$\times$2 unit cell are labeled A--D. Two locations of the LW lattice site, corresponding to right and left trimerizations are shown as filled and open white dots, respectively.  (b) A LW triangular lattice of spacing $2 a_\triangle$ is formed by out-of-plane polarized 1064-nm-wavelength light. Black dots indicate sites of the SW lattice. (C) Combined superlattice potential for the right trimerized kagome lattice.
 (d) Tight binding model of the trimerized kagome lattice, with strong intra-trimer tunneling $J$ and weak inter-trimer tunneling $J'$.}
\label{fig:scheme}
\end{figure*}

In this work, we realize a lattice of triangular plaquettes.  The resulting trimerized (or ``breathing'') kagome lattice has received theoretical interest  \cite{subr95,mila98low,mamb00} as a stepping stone to understanding quantum antiferromagnetism in the kagome lattice, whose ground state properties are not fully known \cite{yan11kagome,iqba14,liao17}.  The trimerized kagome lattice is obtained by choosing one orientation of triangular plaquettes of the kagome lattice to have strong intersite tunnelling ($J$) and letting plaquettes of the other orientation have weak intersite tunnelling ($J'$), as shown in Fig. \ref{fig:scheme} d. For antiferromagnetically coupled spins, the strongly coupled trimers support microscopic spin frustration. The addition of weak inter-trimer coupling leads to robust spin-liquid ground states on macroscopic scales \cite{subr95,mila98low,mamb00}.  

Here, we consider interacting Bose gases within a trimerized kagome lattice with widely tunable trimerization ratio $J/J'$. In the tight-binding limit, our system is modeled by the Hamiltonian
\begin{align}
    &H = - \sum_{\langle p,q \rangle}J_{pq}\left(a_p^\dag a_q + h.c.\right)  + \frac{U}{2} \sum_p n_p(n_p-1) 
    \end{align}

where $\langle p,q \rangle$ denotes summation over all pairs of neighboring lattice sites $p$ and $q$, $J_{pq} = J(J')$ for strongly (weakly) coupled bonds, and $n_p$ is the number operator for site $p$. Here, $J$ and $J' >0$ .

We study the effects of trimerization on the low-energy dynamics and spatial coherence of an ultracold Bose gas of $^{87}$Rb atoms trapped in this lattice.  The spatial inversion asymmetry of the trimerized kagome lattice reveals itself in the momentum distribution of weakly interacting superfluids that are prepared in the trimerized lattice and allowed to undergo transient dynamics after the lattice is perturbed. 

For strongly interacting bosons, the trimerized kagome lattice is predicted to support Mott insulating states in which atoms remain coherently delocalized within trimer plaquettes while inter-trimer coherence is suppressed by interactions \cite{buon05,jurg14dimer,chen16tri}.  We study the spatial coherence of the low-temperature Bose gas in the $U/J' \gg 1$ limit by analyzing the distribution of atoms released from the lattice.  We observe that lattice trimerization causes the nearest-neighbor coherence to remain strong even in the deep Mott insulating limit, long after long-range phase coherence has been lost.  Transiently imprinting site-selective energy shifts on the lattice-trapped gas shows that this spatial coherence resides almost exclusively within the strongly coupled trimer plaquettes.

We form an optical superlattice by overlaying two commensurate triangular lattices, one twice the spacing of the other \cite{jo12kag,thom17scaling}. Each triangular lattice is formed by the intersection of three focused laser beams at equal angles and lying in a single (horizontal) plane \cite{beck10tri}. The short-wavelength (SW) lattice (lattice spacing $a_\triangle = 355$ nm) is formed by \emph{in-plane} polarized, 532-nm-wavelength light, while the long-wavelength (LW) lattice is formed by \emph{out-of plane} polarized 1064-nm-wavelength light. The relative position of these two lattices is stabilized interferometrically to better than 2 nm.

The unit cell of the superlattice contains four sites of the SW lattice (labeled A -- D in Fig.\ \ref{fig:scheme}), and one site of the LW lattice. The trimerized kagome lattice is obtained when the LW lattice site is centered between three equidistant nearest-neighbor sites of the SW lattice, which now form the trimer of the trimerized lattice.  For example, two trimerizations, with opposite spatial inversion asymmetry, are obtained by centering the LW lattice sites within either the A-B-C trimer or the A-C-D trimer. 
The LW lattice has three effects on the overall superlattice.  First, the energies of the three sites in the trimer are lowered relative to the fourth one roughly by $\Delta V \approx \tfrac{1}{2} V_{LW}$ where $V_{LW}$ is the depth of the LW lattice potential.  When $\Delta V$ is sufficiently high, the fourth site is unoccupied and a kagome structure is realized. Second, the LW lattice modifies the tunneling between sites by decreasing the tunneling barrier height between sites within a trimer and increasing the tunneling barrier between trimers. Third, the LW lattice potential causes the combined lattice to ``breathe,'' with the spacing between trimer sites ($a$) decreasing, and the nearest-neighbor spacing between trimers ($a'$) increasing.

We visualize the spatial asymmetry of the trimerized kagome lattice through the coherent diffraction of a superfluid from the lattice. For this, we prepare nearly pure Bose-Einstein condensates of $5\times10^4 \,^{87}$Rb atoms in a hybrid optical and magnetic harmonic trap, with trap frequencies  $(\omega_x, \omega_y, \omega_z) = 2\pi\times(40,70,80)$ Hz, with $\mathbf{z}$ being the vertical axis. The atoms are then loaded into the superlattice by simultaneously increasing depths of the SW and LW lattices to  to $V_{SW}/h= 45$ kHz, $V_{LW}/h = 15$ kHz by an exponential ramp \footnote{The SW and LW lattices are initially turned on at $V_{SW}/h$ = 2 kHz and  $V_{LW}/h$ = 1 kHz respectively to initiate phase stabilization. Both initial lattice depths are sub-recoil and thus have little effect on the atoms. After 20 ms, the depths of the two lattices are simultaneously increased with an exponential ramp. The time constant for the SW lattice ramp is $\tau = 35 $ ms, whilst that for the LW lattice varies between 32 to 58 ms depending on the final depth such that the two ramps always finish at the same time.}. The relative position of the two lattices is adjusted to produce $C_6$ symmetric diffraction patterns of gases released from the lattice.  The gas remains only loosely confined along the transverse (vertical) direction.

\begin{figure}[ht]
\includegraphics{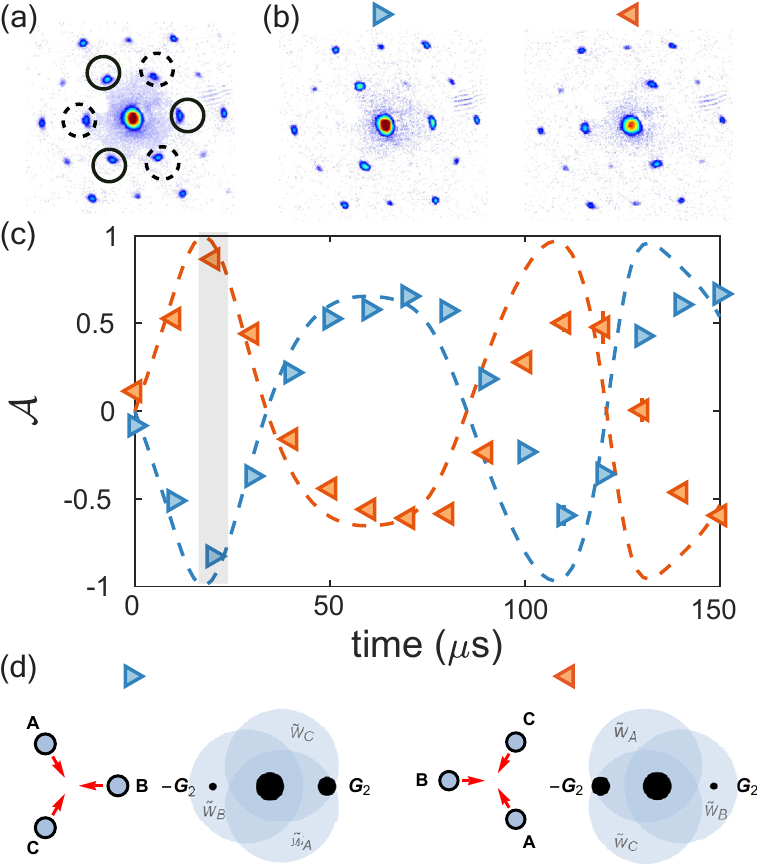}
\caption{Trimerization-dependent momentum space asymmetry. (a)  Momentum distribution of a superfluid within the trimerized kagome lattice shows no asymmetry between diffraction orders at $\mathbf{G}_i$ (solid circles) and $-\mathbf{G}_i$ (dashed circles). (b) Strong inversion asymmetry is seen after $\tau = 20 \, \mu$ s of evolution in the LW lattice potential, with opposite sign for right (blue triangle) and left (red triangle) trimerzations.  (c) Oscillations of the asymmetry parameter $\mathcal{A}$ vs.\ $\tau$ occur with opposite sign for the right and left trimerizations. Each data point represents average over 2-7 iterations. (d) Explanation of the population asymmetry at early $\tau$. Red arrows show the direction of acceleration experienced by the Wannier functions at sites A, B and C for the two different trimerizations. The resultant impulses displace the Wannier functions (contours shown as blue circles) in momentum space. Imbalance between $P_{\mathbf{G}_2}$ and $P_{-\mathbf{G}_2}$ results from interference of these displaced Wannier functions.}
\label{fig:diffraction}
\end{figure}

The momentum space distribution of this superfluid in equilibrium, shown in Fig. \ref{fig:diffraction} a, does not manifest the broken inversion symmetry and reduced rotational symmetry of the lattice, since the superfluid order parameter has uniform phase. To fully characterize the lattice, we imprint complex phases onto the superfluid wavefunction through transient dynamics. After allowing the superfluid 20 ms to equilibrate in the trimerized kagome lattice, we suddenly extinguish the SW lattice potential and allow the atoms to evolve for a variable time $0<\tau<150\,\mu$s in the remaining LW lattice.  We then switch off all potentials, allow the atoms to expand freely for 20 ms, and image their spatial distribution by absorption imaging.

The two trimerization configurations lead to strong but opposite inversion-asymmetric diffraction patterns. We focus on the first order diffraction peaks, occurring at reciprocal lattice vectors $\mathbf{G}_1 = \mathbf{k}_2 - \mathbf{k}_3$ (et cycl.), where $\mathbf{k}_i$ are wavevectors of the LW lattice beams. We observe the diffracted populations $P_{\mathbf{G}_i}$ at wavevectors $\mathbf{G}_i$ to differ from those at $-\mathbf{G}_i$, as shown in Fig.\ \ref{fig:diffraction}(b). We quantify the observed differences by an asymmetry parameter $\mathcal{A} = \left(\sum_i \left(P_{\mathbf{G}_i} - P_{- \mathbf{G}_i}\right) \right)/\left( \sum_i \left(P_{\mathbf{G}_i} + P_{- \mathbf{G}_i}\right)\right)$ \cite{thom16asym}.  We observe equal and opposite oscillations of $\mathcal{A}$ as a function of hold time $\tau$, shown in \ref{fig:diffraction} c, clearly distinguishing the opposite inversion asymmetry of the two trimerization patterns.

To realize the strongly interacting Bose-Hubbard Hamiltonian, we introduce an additional one-dimensional optical lattice, formed by a retroreflected 1064-nm-wavelength light beam propagating along $\mathbf{z}$.   This lattice, with depth $V_\perp/h = 50$ kHz, divides the gas into about 40 layers, each with trap frequencies $(\omega_x, \omega_y, \omega_z ) = 2\pi \times (61,61,22\times10^3)$ Hz. As the tunneling time between layers of 400 ms is slower than the timescale of the experiment, the system can be considered as an ensemble of isolated, two dimensional systems \cite{petr002d}.

The superlattice is ramped up as above, but to variable final lattice depths. The ramp is adiabatic with respect to the band gap, the interaction energy $U$, and the tunneling energies $J$ and $J'$. During the ramp, an additional single-pass, vertically propagating, 1064-nm-wavelength light beam is introduced to provide confinement and maintain a constant Thomas-Fermi radius in each layer.

After being held on for 40 ms, all potentials are simultaneously switched off, and the gas is allowed to expand for 16 ms before being imaged. The observed distribution approximates the momentum distribution of the lattice-trapped gas \cite{gerb08expansion}.

\begin{figure}[t]
\includegraphics{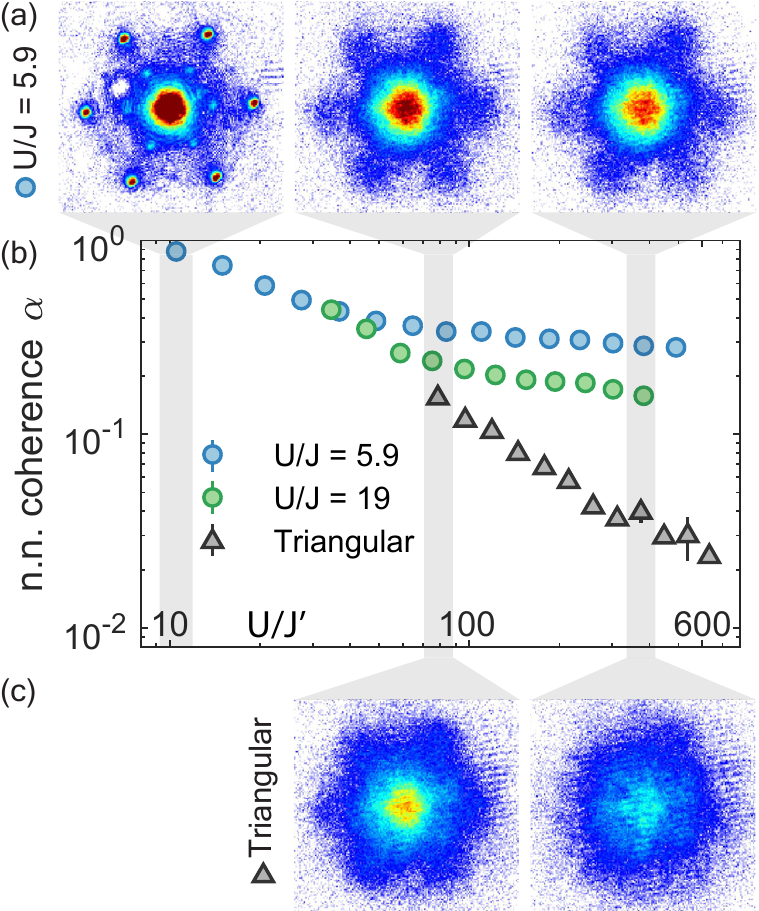}
\caption{Time-of-flight images of strongly interacting atoms released from either the (a) trimerized kagome lattice or (c) triangular lattice. As $U/J'$ is increased at constant $U/J$, sharp diffraction peaks are lost, leaving a broad sixfold symmetric modulation. This modulation persists at large $U/J'$ in the trimerized kagome lattice, but disappears in the triangular lattice. (b) Extracted n.n.\ coherence $\alpha$ for the trimerized kagome (blue, green circles) and triangular lattice (triangles). Data points are the average of 3-9  measurements. Error bars are standard errors of mean.}
\label{fig:persistent}
\end{figure}

In the tight binding limit, the momentum distribution of a lattice-trapped Bose gas is given by 
\begin{equation}
n(\mathbf{k}) = w^*_p(\mathbf{k}) w_q\left(\mathbf{k} \right) \sum_{p,q}^{\mathrm{all}} e^{i\mathbf{k}\cdot(\mathbf{r}_p-\mathbf{r}_q)}\langle b_p^{\dagger}b_q \rangle
\end{equation}

where  $\langle b_p^{\dagger}b_q \rangle$ is the coherence between site $p$ and $q$, with $\mathbf{r}_p$ being the position and $w_p(\mathbf{k})$ is the Fourier-space Wannier function at site $p$, and the summation runs over all lattice sites \cite{gerb08expansion}. When the coherence length of the system is infinite the Fourier sum evaluates to delta functions at the reciprocal lattice vectors, leading to characteristic sharp diffraction peaks observed in the momentum distribution in the superfluid phase. 

For larger $U/J'$ the system is in the Mott insulating regime, and $\langle b_p^{\dagger}b_q \rangle$ vanishes rapidly for distant $p$ and $q$. To leading order we consider only the nearest neighbor (n.n.)\ terms. Then the momentum distribution can be approximated as \cite{gerb05,spie07mott2d, beck10tri}
\begin{equation}
\frac{n(\mathbf{k})}{N} \simeq |\tilde{w}(\mathbf{k})|^2 \bigg( 1 +  \sum_{p<q}^{\text{cell}}\mathrm{Re}\big[ \zeta_{pq} e^{i \mathbf{k} \cdot \mathbf{a}_{pq}} + \zeta'_{pq} e^{i\mathbf{k}\cdot \mathbf{a}'_{pq}} \big] \bigg)
\label{eq:alpha}
\end{equation} 
where $N$ is the total atom number, $\nu$ is the average filling per site, and the indices $p$ and $q$ now run over sites A -- C
in the unit cell. Here, $\textbf{a}_{pq} = \textbf{r}_p - \textbf{r}_q$ is the (intra-trimer)
distance vector between sites $p$ and $q$, and $\zeta_{pq} = \frac{2}{\nu} \langle b_p^{\dagger}b_q \rangle $
quantifies their mutual coherence, evaluated over a bond
with large tunneling energy $J$. Similarly, $a'$, $\zeta'$ are evaluated over an inter-trimer bond with small tunneling energy $J'$. We simplify this expression by assuming an identical cylindrically symmetric Wannier function $\tilde{w}(\textbf{k})$
at each site, neglecting small site-dependent ellipticity at the settings of our experiment.

We measure spatial coherence in the trimerized kagome
lattice at two fixed intra-trimer interaction strengths, $U/J = 5.9$ and $19$ \footnote{Values of $U$, $J$ and $J'$ for each setting of $VSW, VLW$ and
$V\perp$ were calculated following Ref.\cite{walt13abinitio}}. As we increase $U/J^\prime$, the sharply peaked momentum distribution of the superfluid gives way to a broad momentum distribution in the Mott insulating regime ($U/J^\prime \gg 1)$), indicating the loss of long-range phase coherence. 

However, even in the strongly interacting regime, the momentum distribution still shows modulations that indicate the persistence of short-ranged spatial coherence.

We quantify the n.n.\ coherence by fitting the observed distribution with the function
\begin{equation}
\frac{n(\mathbf{k})}{N}\!=\! |\tilde{w}(\mathbf{k})|^2 \bigg( 1 +  \sum_{p<q}^{\text{cell}} \big[ \alpha_{pq} \cos( \mathbf{k}\!\cdot\!\mathbf{a}_{pq})
 + \beta_{pq} \sin(\mathbf{k}\!\cdot\!\mathbf{a}_{pq}) \big] \bigg).
\label{eq:fitting}
\end{equation}

Compared with Eq. 1, this expression is simplified by including only one periodicity in the momentum-space distribution along each lattice direction. This simplification is justified both for weak trimerization, where the lengths $a$ and $a'$ are nearly equal, and also for strong trimerization, where (as we show) the intra-trimer coherence dominates over the inter-trimer coherence. As such, we identify $\alpha_{AB} \simeq \text{Re}(\zeta_{AB} + \zeta'_{BA})$ and $\beta_{AB} \simeq \text{Im}(\zeta_{AB} +\zeta'_{BA})$, and similar for other bond directions. Both the Gaussian width of $\tilde{w}(\textbf{k})$ and also the bond length $a$ are used as fitting parameters and extracted from the images.

For the data of Fig. 3, we set $\beta_{pq} = 0$ since the coherence in this case is expected to be real valued. A single value for the n.n.\ coherence function $\alpha$ is taken as the average of $\alpha_{pq}$ along the three bond directions. Close to the superfluid regime, finer modulations in momentum space, indicating coherence beyond the nearest neighbor, are also present in the images; however, through numerical simulation, we verify that these additional modulations do not affect the fits by which the n.n.\ coherence is extracted.

Our measurements on the trimerized kagome lattice
are benchmarked by additional measurements performed
on atoms in the primitive triangular lattice. For this, we prepare the gas as before, with the exception that the LW lattice light is left off. The coherence function $\alpha$ is measured by the same method described above.

The fitted n.n.\ coherences demonstrate the influence of trimerization of the lattice. At low values of $U/J'$, in the superfluid regime, the n.n.\ coherences of the triangular and trimerized kagome lattices are similar. At larger values of $U/J'$ and
strong trimerization ratio $J/J'$, there is a stark difference. In the triangular lattice, the n.n.\ coherence tends to zero in the Mott insulating limit, scaling as $\alpha \propto (U/J)^{-0.87(9)}$, which is roughly consistent with a perturbative treatment of a Mott insulator with uniform tunneling energies \cite{gerb05,spie07mott2d, beck10tri}. In contrast, for the trimerized kagome lattice n.n.\ coherence remains large due to persistent tunnelling within trimers. We also observe that $\alpha$ is smaller for larger $U/J$, showing the effect of interactions to suppress coherence in a few-site system.

\begin{figure}[t]
\includegraphics{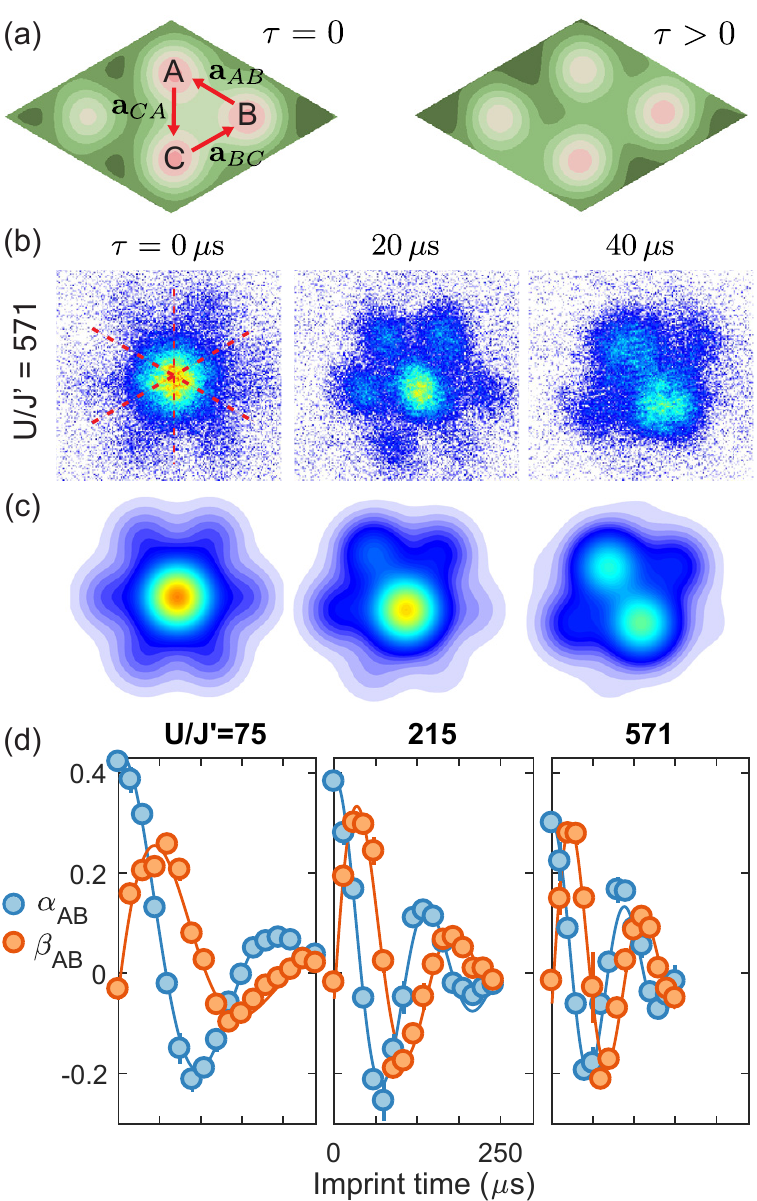}
\caption{Revealing coherence asymmetry via phase imprinting. (A) Change in the trimerized kagome lattice potential when one LW lattice beam is switched off. Site A is detuned from sites B and C by $\Delta V$. (b) The phase imprint from applying this energy offset for variable $\tau$ causes the initially symmetric momentum distribution ($\tau=0$) to evolve. Dashed lines have the same directions as the lattice distance vectors and indicate the directions of modulation.  (c) Qualitatively similar momentum distributions are calculated for a superposition of three identical Gaussian wavefunctions at the sites of a trimer, with a complex phase applied at one site.  The width of the Gaussian function and the visibility of interference are based on experimental parameters. (d) Coherence functions $\alpha_{AB}$ and $\beta_{AB}$ obtained from fitting Eq.\ \ref{eq:fitting} to the observed momentum distributions oscillate out of phase.  The equal amplitudes of oscillation for strong trimerization ($U/J' = 215$ and $571$) show that n.n.\ coherence resides nearly exclusively on the strongly coupled bonds.  Data points are averages of 2-3 measurements. Error bars are standard errors of mean.}
\label{fig:interference}
\end{figure}

The simultaneous lack of long range coherence and persistence of nearest-neighbor coherence implies that the spatial coherence in the trimerized lattice is spatially asymmetric, with large differences between the coherences $\langle b_p^{\dagger}b_q \rangle_J$ and $\langle b_p^{\dagger}b_q \rangle_{J'}$. We demonstrate this fact directly by an interferometric measurement, in which we imprint a site-specific phase on the atomic spatial coherence \cite{taie15lieb}. For this, we turn off one beam of the LW lattice (Fig. \ref{fig:interference}(a)) for a brief time $\tau$, raising the energy of one site in each trimer (A) above the energy of the other two sites (B and C) by an amount $\Delta V \simeq 0.2 \times V_{LW}$. This energy offset causes the coherence functions to become complex, evolving as $\zeta_{AB}(\tau) = e^{i\phi} \zeta_{AB}(0)$ and $\zeta'_{AB}(\tau) = e^{-i\phi} \zeta'_{CA}(0)$, and similar for $\zeta_{AC}$ and $\zeta'_{AC}$, with $\phi = \Delta V \, \tau/\hbar$.

This phase imprint has a pronounced effect on the momentum distribution (Fig.\ \ref{fig:interference}).  In the limit that n.n.\ coherence remains only on the intra-trimer bonds, we expect $\alpha_{AB} \propto \cos(\phi)$ and $\beta_{AB}\propto \sin(\phi)$ to oscillate out of phase and with equal amplitude with $\tau$.  Fitting the observed momentum distribution using Eq.\ \ref{eq:fitting} while allowing for non-zero $\beta_{pq}$ and time-varying spacing $a_{pq}$, we observe such equal amplitude oscillations for the case of strong interactions and strong trimerization, demonstrating that, for these settings, n.n.\ coherence resides nearly exclusively on the intra-trimer bonds.  In the case of weaker trimerization and closer to the superfluid regime, we still observe oscillations in $\beta_{AC}$, but with diminished amplitude with respect to those in $\alpha_{AC}$. This observation demonstrates that, while n.n.\ coherences in this case are still stronger on the intra-trimer bonds, there exists discernible coherence also on the inter-trimer bonds.

The temporal oscillations of $\alpha(\tau)$ ($\beta(\tau)$) are not purely cosinusoidal (sinusoidal), and appear to decay in time.  This feature may be explained by residual effects of interactions and tunneling after the lattice modification.  A three-site Bose-Hubbard model calculation, accounting for the interaction energy $U$, tunneling energy $J$ and energy offset $\Delta V$ within a single trimer, reproduces the observed behavior as resulting from the few-body energy spectrum, up to an overall scaling of the coherence \footnote{See Supplemental Material at [URL will be inserted by publisher] for description of Bose-Hubbard model calculations}.

The trimerized kagome lattice presents a new setting for experimental quantum simulation of condensed-matter and many-body quantum physics. Our present work is performed on gases with a high filling factor, reaching as high as $\nu = 8/3$ atoms per site (8 atoms per trimer) at the center of the gas. Reaching filling factors below $\nu=1$ would allow for definitive studies of the predicted fractional (per site) Mott insulating state \cite{jurg14dimer}.
Future experiments may identify few-body eigenstates within single trimers can be identified by precise spectroscopy.  These eigenstates include circulating states, to which atoms can be driven to to realize quantum simulations of materials with orbital magnetism  \cite{dams05trifermi}.  Alternately, trapping two-state fermions within the trimerized kagome lattice at half filling can simulate the spin-1/2 Heisenberg antiferromagnet, which is expected to have a spin-liquid ground state \cite{sant04kagome, dams05trimerized, dams05trifermi}.

We thank Y.M.\ Tso for experimental assistance. This work was supported by the NSF, and by the AFOSR and ARO through the MURI program (grant numbers FA9550-14-1-0035 and W911NF-17-1-0323, respectively).


\begin{thebibliography}{10}

\bibitem{trot08superexchange}
S.~Trotzky, P.~Cheinet, S.~Folling, M.~Feld, U.~Schnorrberger, A.~M. Rey,
  A.~Polkovnikov, E.~A. Demler, M.~D. Lukin, and I.~Bloch.
\newblock {Time-resolved observation and control of superexchange interactions
  with ultracold atoms in optical lattices}.
\newblock {\em Science}, 319:295--299, 2008.

\bibitem{nasc12rvb}
S.~Nascimb\`{e}ne, Y.~A. Chen, M.~Atala, M.~Aidelsburger, S.~Trotzky,
  B.~Paredes, and I.~Bloch.
\newblock {Experimental Realization of Plaquette Resonating Valence-Bond States
  with Ultracold Atoms in Optical Superlattices}.
\newblock {\em Phys. Rev. Lett.}, 108:205301, 2012.

\bibitem{grei13}
Daniel Greif, Thomas Uehlinger, Gregor Jotzu, Leticia Tarruell, and Tilman
  Esslinger.
\newblock {Short-Range Quantum Magnetism of Ultracold Fermions in an Optical
  Lattice}.
\newblock {\em Science}, 340:1307--1310, 2013.

\bibitem{dai17toric}
Han-Ning Dai, Bing Yang, Andreas Reingruber, Hui Sun, Xiao-Fan Xu, Yu-Ao Chen,
  Zhen-Sheng Yuan, and Jian-Wei Pan.
\newblock {Four-body ring-exchange interactions and anyonic statistics within a
  minimal toric-code Hamiltonian}.
\newblock {\em Nature Physics}, 13:1195, 2017.

\bibitem{subr95}
V.~Subrahmanyam.
\newblock {Block spins and chirality in the frustrated Heisenberg model on
  kagom$\acute{e}$ and triangular lattices}.
\newblock {\em Phys. Rev. B}, 52:1133--1137, 1995.

\bibitem{mila98low}
F.~Mila.
\newblock {Low-Energy Sector of the
  $\mathit{S}\mathit{}\phantom{\rule{0ex}{0ex}}=\phantom{\rule{0ex}{0ex}}1/2$
  Kagome Antiferromagnet}.
\newblock {\em Phys. Rev. Lett.}, 81:2356--2359, 1998.

\bibitem{mamb00}
M.~Mambrini and F.~Mila.
\newblock {RVB description of the low-energy singlets of the spin 1/2 kagomé
  antiferromagnet}.
\newblock {\em Eur. Phys. J. B}, 17:651--659, 2000.

\bibitem{yan11kagome}
Simeng Yan, David~A. Huse, and Steven~R. White.
\newblock {Spin-Liquid Ground State of the S = 1/2 Kagome Heisenberg
  Antiferromagnet}.
\newblock {\em Science}, 332:1173--1176, 2011.

\bibitem{iqba14}
Yasir Iqbal, Didier Poilblanc, and Federico Becca.
\newblock {Vanishing spin gap in a competing spin-liquid phase in the kagome
  Heisenberg antiferromagnet}.
\newblock {\em Phys. Rev. B}, 89:020407, 2014.

\bibitem{liao17}
H.~ J Liao, Z.~ Y Xie, J.~Chen, Z.~ Y Liu, H.~ D Xie, R.~ Z Huang,
  B.~Normand, and T.~Xiang.
\newblock {Gapless Spin-Liquid Ground State in the $S=1/2$ Kagome
  Antiferromagnet}.
\newblock {\em Phys. Rev. Lett.}, 118:137202, 2017.

\bibitem{buon05}
P.~Buonsante, V.~Penna, and A.~Vezzani.
\newblock {Fractional-filling Mott domains in two-dimensional optical
  superlattices}.
\newblock {\em Phys. Rev. A}, 72:031602, 2005.

\bibitem{jurg14dimer}
Ole J\"{u}rgensen and Dirk-Sören L\"{u}hmann.
\newblock {Dimerized Mott insulators in hexagonal optical lattices}.
\newblock {\em New Journal of Physics}, 16:093023, 2014.

\bibitem{chen16tri}
Qi-Hui Chen, Peng Li, and Haibin Su.
\newblock {Fractional Mott insulator-to-superfluid transition of Bose–Hubbard
  model in a trimerized Kagomé optical lattice}.
\newblock {\em J. Phys: Cond. Matter}, 28:256001, 2016.

\bibitem{jo12kag}
Gyu-Boong Jo, Jennie Guzman, Claire~K. Thomas, Pavan Hosur, Ashvin Vishwanath,
  and Dan~M. Stamper-Kurn.
\newblock {Ultracold Atoms in a Tunable Optical Kagome Lattice}.
\newblock {\em Phys. Rev. Lett.}, 108:045305, 2012.

\bibitem{thom17scaling}
Claire~K. Thomas, Thomas~H. Barter, Tsz-Him Leung, Masayuki Okano, Gyu-Boong
  Jo, Jennie Guzman, Itamar Kimchi, Ashvin Vishwanath, and Dan~M. Stamper-Kurn.
\newblock {Mean-Field Scaling of the Superfluid to Mott Insulator Transition in
  a 2D Optical Superlattice}.
\newblock {\em Phys. Rev. Lett.}, 119:100402, 2017.

\bibitem{beck10tri}
C.~Becker, P.~Soltan-Panahi, J.~Kronjager, S.~Doscher, K.~Bongs, and
  K.~Sengstock.
\newblock {Ultracold quantum gases in triangular optical lattices}.
\newblock {\em New Journal of Physics}, 12:065025, 2010.

\bibitem{Note1}
The SW and LW lattices are initially turned on at $V_{SW}/h$ = 2 kHz and
  $V_{LW}/h$ = 1 kHz respectively to initiate phase stabilization. Both initial
  lattice depths are sub-recoil and thus have little effect on the atoms. After
  20 ms, the depths of the two lattices are simultaneously increased with an
  exponential ramp. The time constant for the SW lattice ramp is $\tau = 35 $
  ms, whilst that for the LW lattice varies between 32 to 58 ms depending on
  the final depth such that the two ramps always finish at the same time.

\bibitem{thom16asym}
C.~K. Thomas, T.~H. Barter, T.-H. Leung, S.~Daiss, and D.~M. Stamper-Kurn.
\newblock {Signatures of spatial inversion asymmetry of an optical lattice
  observed in matter-wave diffraction}.
\newblock {\em Phys. Rev. A}, 93:063613, 2016.

\bibitem{petr002d}
D.~S. Petrov, M.~Holzmann, and G.~V. Shlyapnikov.
\newblock {Bose-Einstein Condensation in Quasi-2D Trapped Gases}.
\newblock {\em Phys. Rev. Lett.}, 84:2551--2555, 2000.

\bibitem{gerb08expansion}
F.~Gerbier, S.~Trotzky, S.~Fölling, U.~Schnorrberger, J.~D. Thompson,
  A.~Widera, I.~Bloch, L.~Pollet, M.~Troyer, B.~Capogrosso-Sansone, N.~V.
  Prokof’ev, and B.~V. Svistunov.
\newblock {Expansion of a Quantum Gas Released from an Optical Lattice}.
\newblock {\em Phys. Rev. Lett.}, 101:155303, 2008.

\bibitem{gerb05}
Fabrice Gerbier, Artur Widera, Simon Folling, Olaf Mandel, Tatjana Gericke, and
  Immanuel Bloch.
\newblock {Phase Coherence of an Atomic Mott Insulator}.
\newblock {\em Phys. Rev. Lett.}, 95:050404--4, 2005.

\bibitem{spie07mott2d}
I.~B. Spielman, W.~D. Phillips, and J.~V. Porto.
\newblock {Mott-Insulator Transition in a Two-Dimensional Atomic Bose Gas}.
\newblock {\em Phys. Rev. Lett.}, 98:080404, 2007.

\bibitem{Note2}
Values of $U$, $J$ and $J'$ for each setting of $VSW, VLW$ and $V\perp $ were
  calculated following Ref.\cite {walt13abinitio}.

\bibitem{taie15lieb}
Shintaro Taie, Hideki Ozawa, Tomohiro Ichinose, Takuei Nishio, Shuta Nakajima,
  and Yoshiro Takahashi.
\newblock {Coherent driving and freezing of bosonic matter wave in an optical
  Lieb lattice}.
\newblock {\em Science Advances}, 1, 2015.

\bibitem{Note3}
See Supplemental Material at [URL will be inserted by publisher] for
  description of Bose-Hubbard model calculations.

\bibitem{dams05trifermi}
B.~Damski, H.~U. Everts, A.~Honecker, H.~Fehrmann, L.~Santos, and
  M.~Lewenstein.
\newblock {Atomic Fermi Gas in the Trimerized Kagom\'e Lattice at $2/3$
  Filling}.
\newblock {\em Phys. Rev. Lett.}, 95:060403, 2005.

\bibitem{sant04kagome}
L.~Santos, M.A. Baranov, J.L. Cirae, H.-U. Everts, H.~Fehrmann, and
  M.~Lewenstein.
\newblock {Atomic Quantum Gases in Kagome Lattices}.
\newblock {\em Phys. Rev. Lett.}, 93:030601, 2004.

\bibitem{dams05trimerized}
B.~Damski, H.~Fehrmann, H.~U. Everts, M.~Baranov, L.~Santos, and M.~Lewenstein.
\newblock {Quantum gases in trimerized kagom\'{e} lattices}.
\newblock {\em Phys. Rev. A}, 72:053612, 2005.

\end{thebibliography}

\onecolumngrid
\newpage

\section*{Supplementary Material}
\beginsupplement

\section{Bose-Hubbard Calculation}

In our explanation of the phase-imprint technique in the text, we referred to a three site Bose-Hubbard model calculation. In this section we provide further details of both this calculation, and its explanation of the observed temporal decay of the coherences $\alpha_{AB}$ and $\beta_{AB}$ in the phase imprint technique.

The calculation begins by exactly diagonalizing the three site Bose-Hubbard model $H_I$ below in the $N$ particle sector,
\begin{equation}
H_{I}=-J\sum_{<p,q>}b_{p}^{\dagger}b_{q}+\frac{U}{2}\sum_{p=A,B,C}n_{p}(n_{p}-1),
\end{equation}
where $n_p$ is the number of atoms on site $p \in \{A,B,C\}$.

We then diagonalize the same Hamiltonian with an energy offset $\Delta V$ on site A:

\begin{equation}
H_{PI}=H_{I}+\Delta V n_{A}.
\end{equation}

We project the ground state of $H_I$ onto the eigenstates of $H_{PI}$, evolve eigenstates at eigenfrequencies and construct the nearest-neighbor coherences,

\begin{eqnarray}
\alpha_{AB} &=& \frac{2}{v} \mathrm{Re}(b_{A}^{\dagger}b_{B}) \\
\beta_{AB} &=& \frac{2}{v} \mathrm{Im}(b_{A}^{\dagger}b_{B}).
\end{eqnarray}

We do not account for changes to $U$ and $J$ in the new potential. We also neglect occupation of the 4th site and tunneling between plaquettes, which is appropriate only for short timescales.

\subsection{Data Comparison}

\begin{figure}[t]
    \centering
    \includegraphics{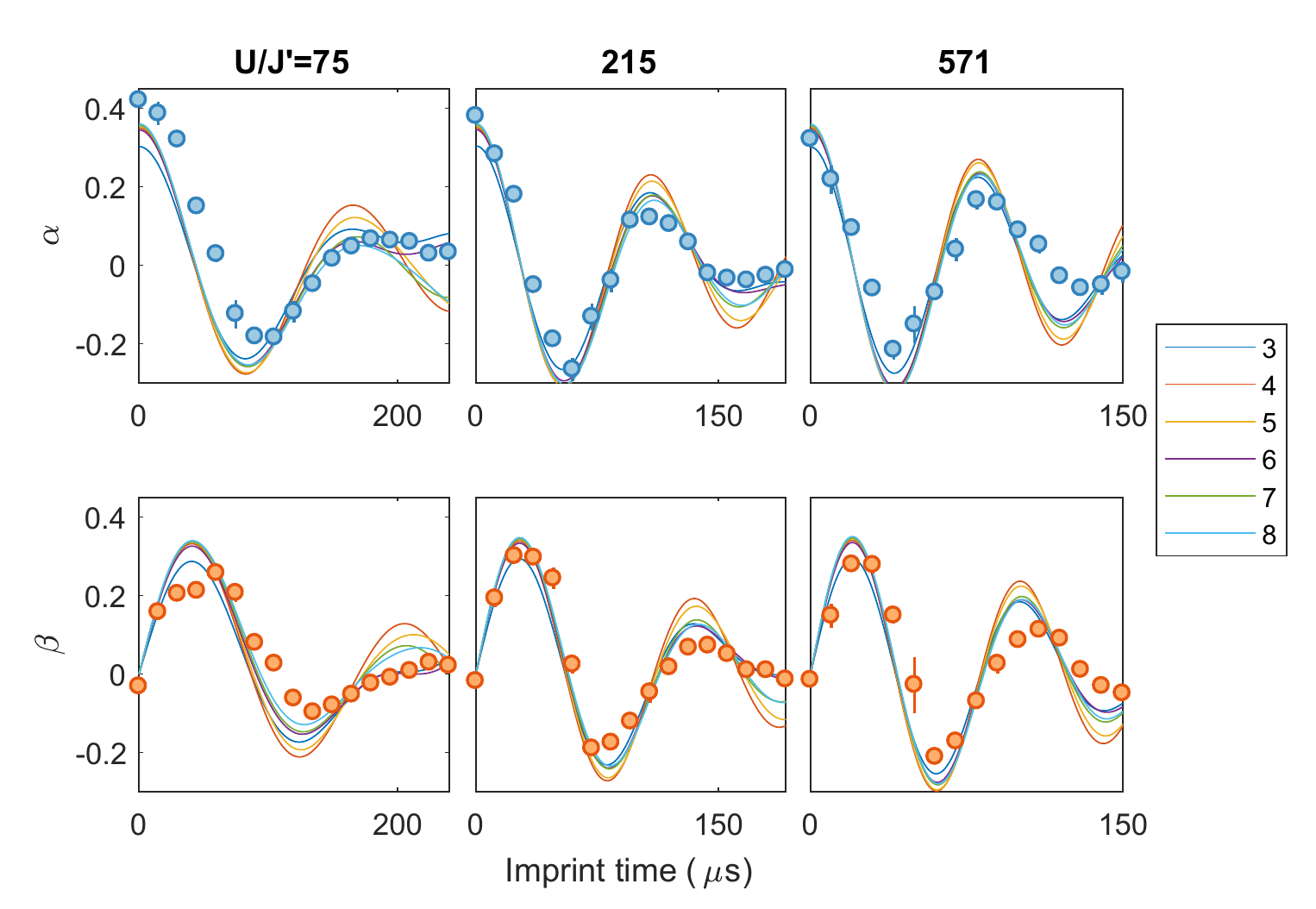}
    \caption{Comparison of the nearest neighbor coherence data (circles) to an exact solution of the three site Bose-Hubbard model (solid lines) for the phase imprint experiment. Theory lines have been uniformly scaled by 2/3 in the y-axis. The multiple theory lines correspond to different particle number $\nu$ per trimer.}
    \label{fig:BH_PhaseImprintl}
\end{figure}

We find good agreement between the coherences $\alpha_{AB}$ and $\beta_{AB}$ calculated from this model and those obtained in data for a range of particles per trimer, as shown in figure \ref{fig:BH_PhaseImprintl}. We note that all of the theoretical coherences have been scaled by a constant factor of 2/3 to obtain a good agreement. We believe that this constant scaling accounts for the non-zero temperature of the data. The decay of both the $\alpha_{AB}$ and $\beta_{AB}$ coherences is clearly visible in the theory.

\subsection{Explanation}
\begin{figure}[t]
    \centering
    \includegraphics{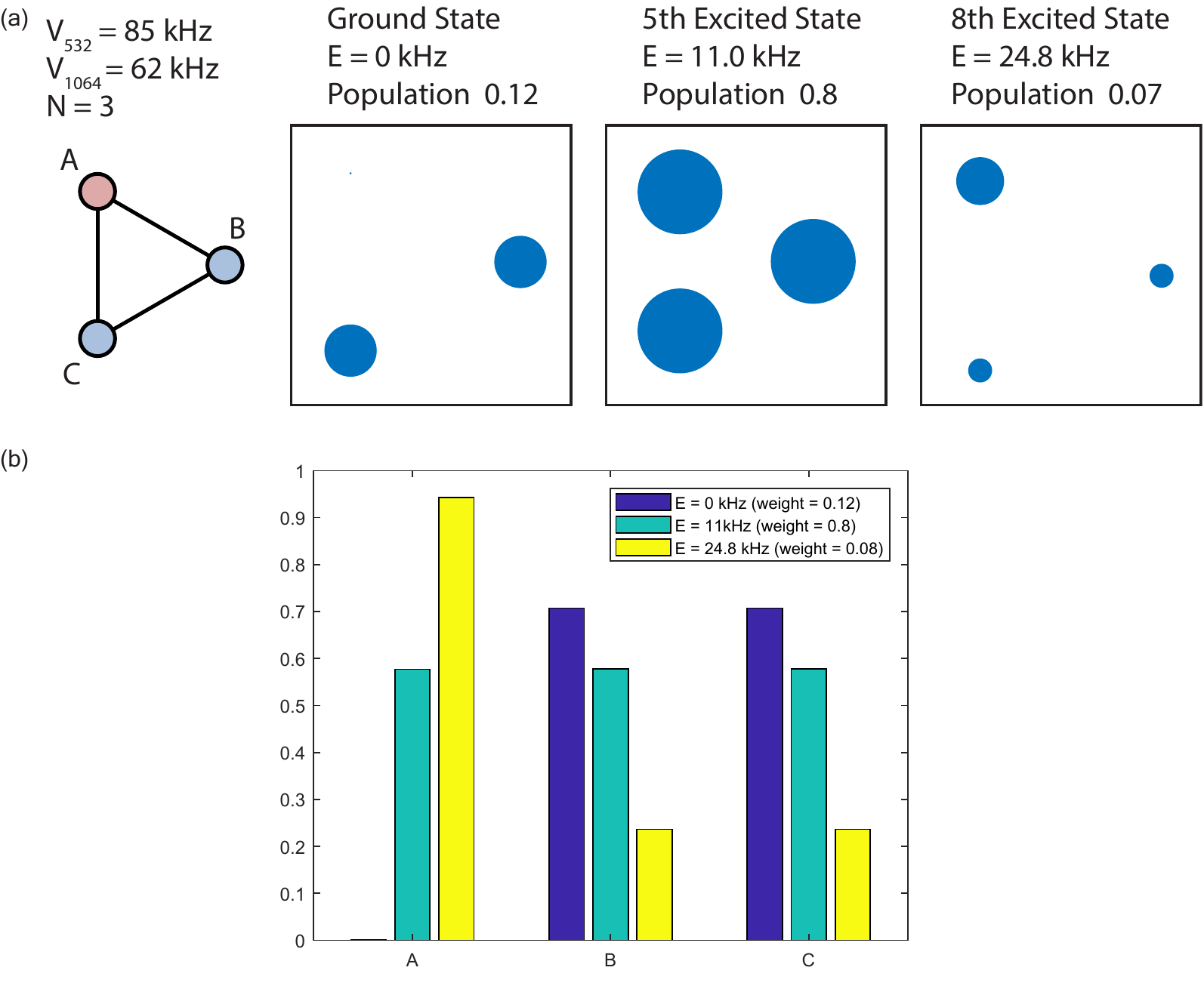}
    \caption{
(a) Densities of the eigenstates responsible for the beating in the phase imprint experiment. The ground band of the trimerized kagome lattice
at $V_{SW}/h=85$ kHz, $V_{LW}/h=62$ kHz ($U = $1.7 kHz, $J =$ 3 Hz) mostly projects into three eigenstates of the new potential. The populations of these eigenstates in the three sites A, B and C are depicted by the areas of the blue circles. The largest population is an equal density state (5th excited state) but a significant population exists in the ground state and the eighth excited states, all of which have different interaction energies. Interference occurs between the ground and 5th excited state at 11.0 kHz, and between the fifth and eight excited
states at 13.8 kHz. (b) The probability density at the three sites of a trimer for the three most highly populated engenstates in the new basis.}
    \label{fig:BH_Explanation}
\end{figure}

To illuminate the cause of this decay, we focus on the $N=3$ sector. The ground state of the initial Hamiltonian $H_I$ has equal population in the sites A, B and C. This ground state is primarily projected into three eigenstates of $H_{PI}$, which are shown in figure \ref{fig:BH_Explanation}. The most populated state is an excited state with energy $\Delta V+O(U)$ but the ground state of energy $O(U)$ and an excited state of energy $2 \Delta V+O(U)$ are also substantially populated. Interference between the most populated state and the other two states leads to an oscillation in the coherence at approximately $\Delta V$, but because the interaction energies of all three of these states are different, we see a two frequency oscillation. Over the short section of data taken this two frequency evolution is not well resolved and looks like decay.

\end{document}